\documentclass[aps,pra,twocolumn,superscriptaddress,showpacs]{revtex4}

\usepackage{epsfig}

\begin{document}
\title{
Single-cell atomic quantum memory for light
}
\author{Tom\'{a}\v{s} Opatrn\'{y}}
\affiliation{Department of Theoretical Physics, Palack\'{y} University, 
17. listopadu 50, 77200 Olomouc, Czech Republic}


\date{\today}

\begin{abstract}

Recent experiments demonstrating atomic quantum memory for light [B. Julsgaard
et al., Nature {\bf 432,} 482 (2004)] involve two macroscopic samples of atoms,
each with opposite spin polarization. It is shown here that a single atomic cell
is enough for the memory function if the atoms are optically pumped with
suitable linearly polarized light, and  quadratic Zeeman shift and/or ac Stark
shift are used to manipulate rotations of the quadratures. This should  enhance
the  performance of our quantum memory devices since less resources are needed
and losses of light in crossing different media boundaries are avoided.

\end{abstract}


\maketitle


\section{Introduction}

Recently, quantum memory for light has been demonstrated in the Copenhagen lab
\cite{FiurasekNature04} where the setup was almost identical to a previous
experiment demonstrating long-lived entanglement of two macroscopic objects
\cite{JulsgaardNature01}. Both schemes involve two cells with macroscopic
numbers  of cesium atoms whose spins are polarized in antiparallel directions,
perpendicular to the direction of light propagation through the cells.
Off-resonant interaction between the beam and the atoms causes ac Stark shift of
the atomic levels and Faraday rotation of the light polarization. It has been
shown that the combined effect corresponds to the quantum nondemolition (QND)
interaction between atomic and light variables \cite{Kuzmich98}  which can be
used in many quantum information protocols \cite{Kuzmich00}. In both schemes 
\cite{FiurasekNature04,JulsgaardNature01} the two samples are placed in a
homogeneous magnetic field which causes Larmor precession of the atomic spins.
The reason for this is that the polarization rotation is then observed at frequency  sidebands sufficiently
displaced from the light frequency $\omega_0$,   which enables us to avoid 
technical noises.
Whereas in the entanglement scheme \cite{JulsgaardNature01} the presence of two
samples was essential to demonstrate entanglement of two objects, in the memory
scheme  \cite{FiurasekNature04} it just helps constructing the QND Hamiltonian
with the precessing spins: two light modes oscillating at frequencies $\omega_0
+ \Omega$ and  $\omega_0 - \Omega$ must be somehow matched with two atomic
modes, and using two samples does the job. (Were there no technical noises, one could use directly the carrier optical frequency $\omega_0$ and work with a single memory cell without using magnetic field and Larmor precession.)

Although the scheme demonstrates the
feasibility of a memory preserving quantum features of light pulses, it has some
disadvantages. First, using two vapor cells per each stored mode  increases the
resources needed. Moreover, each time the light beam crosses the walls of the cell,
losses occur which increase the noise. This could lead to distortion of some
quantum features that then could not be recovered with sufficient fidelity. We
propose here a scheme that uses a single vapor cell and still is able to involve
the full QND interaction between atoms and optical signal at a sideband mode.
The working medium and physical parameters  chosen for the discussion are
similar to those  in the Copenhagen experiment \cite{FiurasekNature04}, even
though the principles are valid for more general atomic systems. The important
tricks are initial pumping of the atoms with a suitable linearly polarized
light, and applying magnetic, optical, or microwave pulses to induce nonlinear Zeeman or Stark splitting of the atomic levels which transform the relevant quadratures.

\begin{figure}
\centerline{\epsfig{file=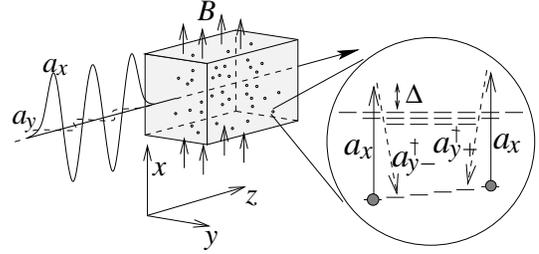,scale=0.4}}
\caption{\label{f-scheme1}
Geometry of the setup and scheme of the interaction in $x$ quantization.
Light pulse travels in the $z$ direction, 
the $x$ polarization
mode is in a strong coherent state and the $y$ polarization is weakly
excited, carrying quantum signal. The pulse goes through atomic vapor placed in
magnetic field pointing in the $x$ direction, the light being detuned from
resonance of some electric dipole transition. The inset scheme illustrates
the Raman transitions described by Hamiltonian (\ref{hAFprime}).
}
\end{figure}


\section{Atom-light interaction}

Let us first discuss the principles of the 
atom-light QND interaction (see also
\cite{OpatrnyFiurasek05}). Atoms in the cell interact with light pulses
traveling in the $z$ direction  and with a magnetic field in the $x$ direction
(see Fig. \ref{f-scheme1}).  
The atoms are initially pumped into one of the hyperfine levels of the ground
electronic state.
 The magnetic field $B$ causes Larmor precession
of the atomic spins around the $x$ axis with frequency $\Omega \propto B$.
Each pulse has a strong
component which is $x$ polarized and oscillates at 
optical frequency $\omega_0=2\pi c/\lambda$ 
and a weak quantum component which is
$y$-polarized. 
The atom-field interaction Hamiltonian is 
\begin{equation}
  H_{AF} = -i\hbar
 \sum_{m} G_m a_x
 \left( a_{y+}^{\dag}\sigma_{m,m+1} + a_{y-}^{\dag}
 \sigma_{m+1,m} \right) + h.c.
 \label{hAFprime}
\end{equation}
Here $a_x$ is the annihilation operator of the $x$ polarized photons at
frequency $\omega_0$ off resonance of some atomic transition line, 
$a_{y+}^{\dag}$ and $a_{y-}^{\dag}$ are the creation operators
of $y$ polarized photons at the sideband frequencies $\omega_0+\Omega$
and  $\omega_0-\Omega$, respectively,
$\sigma_{m,m'}=|m \rangle \langle m'|$ is the coherence between the atomic
magnetic states $m$ and $m'$ in the relevant hyperfine level
with $x$ as the quantization axis. 
The coupling constant $G_m$ depends on the used isotope 
and particular transition. As an example, working with cesium 
pumped into the $F=4$ hyperfine level of the ground electronic state
$6^2S_{1/2}$
and using light near the D2 line, the coupling is
$G_m=\mu_0^2 E_0^2/(48 \hbar^2 \Delta)\sqrt{20-m(m+1)}$,
where  $\mu_0$ is the dipole moment element of the optical transition
related to the spontaneous decay rate $\gamma$ by
$\mu_0^2=3\epsilon_0 \hbar \lambda^3 \gamma/(2\pi^2)$,
$E_0$ is
the vacuum electric field, $E_0^2=\hbar \omega_0/(2 \epsilon_0 V)$,  $V=AcT$ is
the quantization volume, $A$ and $T$  are the transversal area and duration of
the optical pulse, respectively. The detuning $\Delta$ is the frequency
difference between the optical field and the given atomic D2 transition 
and is assumed to be much larger than the hyperfine 
splitting in
that level and much smaller than detuning from any other atomic level.
It is convenient to work with nonmonochromatic
modes $a_{yC}=2^{-1/2} (a_{y-}+a_{y+})$ 
and $a_{yS}=2^{-1/2}(a_{y-}-a_{y+})$,
whose field quadratures $X_j = 2^{-1/2}(a_j + a^{\dag}_j)$, $P_j =
-i2^{-1/2}(a_j-a^{\dag}_j)$ are measured using homodyne detection as the cosine
and sine signal components oscillating at frequency $\Omega$. 
The Hamiltonian (\ref{hAFprime}) is simplified
when the
$a_x$ mode is in a strong coherent state $|\alpha_0\rangle$ with $\alpha_0$
real and the atoms are initially prepared in 
one of the extreme states,
$|F,m=-F\rangle$ or $|F,m=F\rangle$. Let us assume that atoms
denoted by index 1 are
initially pumped into state $|F,m=-F\rangle$
so that
only the coherences $\sigma_{-F,m}$ and  $\sigma_{m,-F}$ are
non-negligible,   
and atoms denoted by index 2 are
initially pumped into state $|F,m=F\rangle$
so that
only the coherences $\sigma_{F,m}$ and  $\sigma_{m,F}$ are
non-negligible.
Then after summing over all $N_A$ atoms of each kind, the interaction
Hamiltonians for the two classes of atoms  
become
\begin{eqnarray}
 H_{\rm int}^{(1)} &=& \hbar \kappa ( P_C X_{A1} + X_S P_{A1}) ,
 \label{QND1} \\
 H_{\rm int}^{(2)} &=& \hbar \kappa ( P_C X_{A2} - X_S P_{A2}) .
 \label{QND2}
\end{eqnarray}
Here
the atomic quadratures are defined as
\begin{eqnarray}
 X_{A1} &=& \frac{1}{\sqrt{2 N_A}} \sum_{k=1}^{N_A}(
\sigma^{(k)}_{-F,-F+1} + \sigma^{(k)}_{-F+1,-F}), \\
 P_{A1} &=& \frac{-i}{\sqrt{2 N_A}}\sum_{k=1}^{N_A}(
\sigma^{(k)}_{-F,-F+1} - \sigma^{(k)}_{-F+1,-F}), \\
 \label{XA2}
 X_{A2} &=& \frac{1}{\sqrt{2 N_A}} \sum_{k=1}^{N_A}(
\sigma^{(k)}_{F-1,F} + \sigma^{(k)}_{F,F-1}), \\
 \label{PA2}
 P_{A2} &=& \frac{i}{\sqrt{2 N_A}}\sum_{k=1}^{N_A}(
\sigma^{(k)}_{F-1,F} - \sigma^{(k)}_{F,F-1}), 
\end{eqnarray}
the coupling constant is $\kappa = -E_0^2\mu_0^2 \sqrt{N_L N_A}
/(12\hbar^2 \Delta)$,
with the photon number $N_L=|\alpha_0|^2$, and the index $k$ denoting individual
atoms. Since for atoms 1 the populations $\sigma^{(k)}_{-F,-F} \approx 1$, and
$\sigma^{(k)}_{m,m} \approx 0$ for $m\neq -F$, one can see that
$X_{A1}$ and $P_{A1}$ satisfy the commutation relation $[X_{A1},P_{A1}]=i$,
and similarly  for atoms 2, $[X_{A2},P_{A2}]=i$.

If the light beam interacts with both classes of atoms, the total Hamiltonian
$H_{\rm int}=H_{\rm int}^{(1)}+H_{\rm int}^{(2)}$ can be written as
\begin{eqnarray}
 H_{\rm int} &=& \sqrt{2} \hbar \kappa ( P_C X_{A+} + X_S P_{A-}) ,
 \label{QNDtot} 
\end{eqnarray}
where 
\begin{eqnarray}
 X_{A\pm} &=& \frac{1}{\sqrt{2}} (X_{A1}\pm X_{A2}), \\
 P_{A\pm} &=& \frac{1}{\sqrt{2}} (P_{A1}\pm P_{A2}) ,
\end{eqnarray}
and the quadratures satisfy the cannonical commutation relations $[X_{A+},P_{A+}]=i$ and $[X_{A-},P_{A-}]=i$.
The Hamiltonian (\ref{QNDtot}) describes the QND interactions 
in two independent pairs
of systems. The atomic ``$+$'' mode interacts with the light cosine mode and
the atomic ``$-$'' mode interacts with the light sine mode. In contrast, each
of the Hamiltonians (\ref{QND1}) and  (\ref{QND2}) would lead to 
unwanted intermodal
coupling of light modes C and S. Thus, with precessing spins
one needs two classes of atoms to
construct
a QND interaction coupling one atomic mode with one light mode.

In the experiment \cite{FiurasekNature04} these two classes of oppositely
polarized atoms were kept separately in two vapor cells. The initial pumping
was achieved by circularly polarized light beams propagating in the $x$
direction. However, there is no principal reason why these two classes of atoms
should not share the same cell. One only has to deal with a few tasks: pump the
atoms into a mixture of states $|F,m=\pm F\rangle$, make sure that the
coherences in atoms 1 and 2 oscillate with the same frequencies, and make sure
that one can rotate the atomic quadratures of each mode on demand.


\begin{figure}
\centerline{\epsfig{file=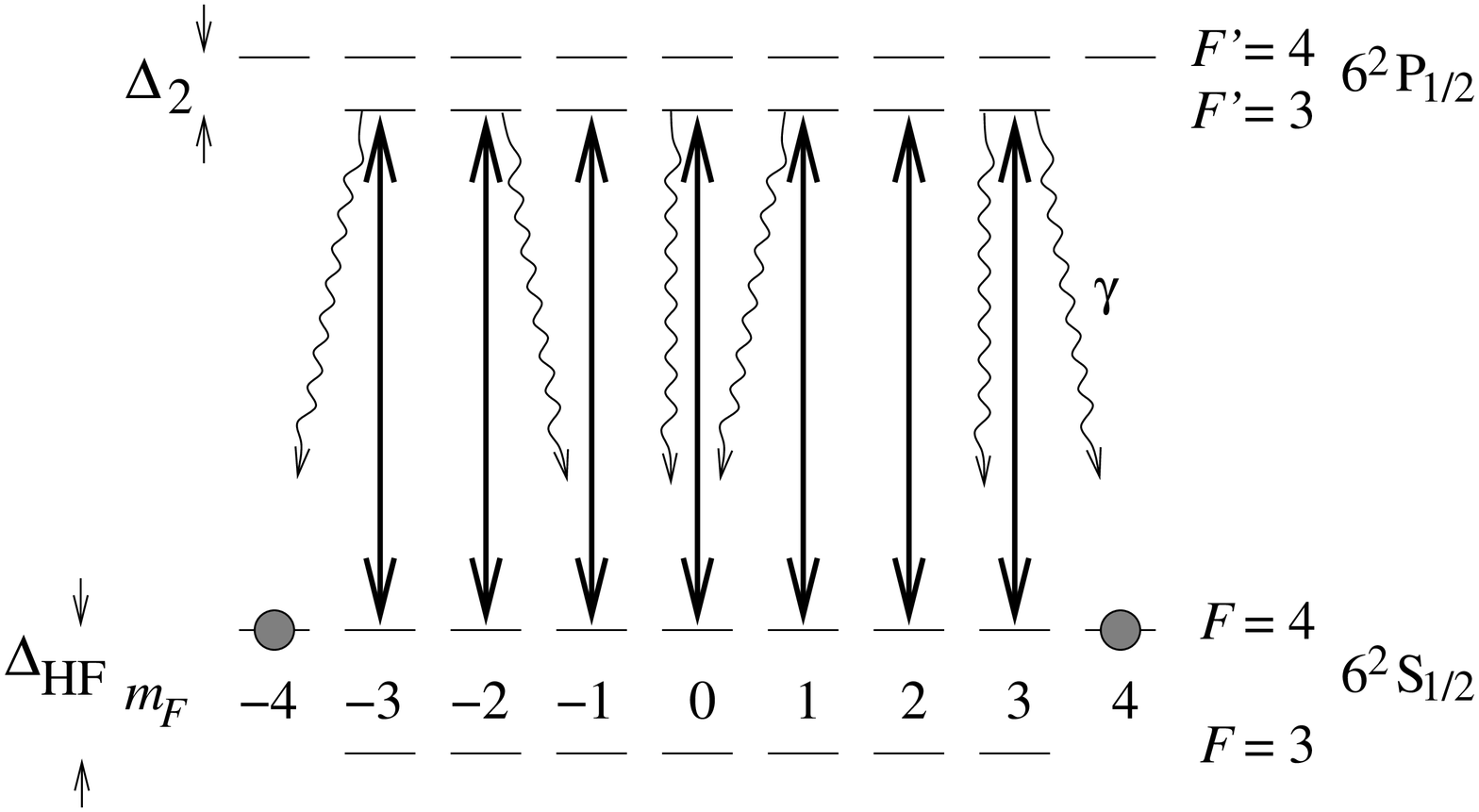,scale=0.3}}
\caption{\label{f-pumping1}
Scheme of the pumping process in cesium
in $x$-quantization. The pump beam is linearly
polarized in the $x$-direction and is resonant with the atomic transition
$F=4 \leftrightarrow F'=3$. 
The most extreme states
$|F=4,m_F=\pm 4\rangle$ are dark with respect to the pump. Repump field (not
shown) drives the atoms out of the manifold $F=3$ so that after some time each
atom ends up in one of the two dark states.
}
\end{figure}


\section{Pumping}

To prepare a mixture of both atomic classes one can use a light beam
propagating in the $z$ direction, linearly polarized in the $x$ direction and
tuned into resonance with the $F \leftrightarrow F'=F-1$ transition (see Fig.
\ref{f-pumping1}). Light polarized along the quantization axis couples states
with the same magnetic quantum number $m$ so that the two extreme states  with
$m=\pm F$ remain uncoupled and act as dark states with respect to the pump
field. If one applies also a repump field driving the atoms out of the
remaining hyperfine level  (similarly as in
\cite{FiurasekNature04,JulsgaardNature01}), all atoms finally end up in one of
the two dark states. With cesium atoms, one can drive the transition between
the electronic ground state 6$^2$S$_{1/2},F=4$ and the excited state 
6$^2$P$_{1/2},F'=3$ with the D1 line 894.6 nm light. 

To avoid coupling of the
pumping beam to the 
$m=\pm F$ states, one must make sure that the pump light stays sufficiently far
from resonance of the neighboring 6$^2$P$_{1/2},F'=4$ state. 
This means that one has to take into account Doppler broadening of the
transition. Fortunately, the Doppler linewidth of the room temperature cesium
atoms is $\Delta_{\rm Doppler}\sim$190~MHz \cite{JulsgaardThesis03} which is much less than the
hyperfine splitting of the D1 line $\Delta_2=$ 1168~MHz. The occupation probability of the unwanted states $m\neq \pm F$ is proportional to 
$\gamma \Delta_{\rm Doppler}/\Delta_2^2 \approx 10^{-3}$, which is sufficiently low.

A more serious effect can stem from the fact 
that during collisions of oppositely polarized atoms the electrons can exchange their spins. 
During such a collision the atom can be brougth out of the manifold of states with $F=4$, and/or, the correlations of its coherences $\sigma_{m,m'}$ with other atoms in the sample are averaged to zero.
This effect (absent in samples with all spins oriented in the same direction) will contribute to the decoherence. Fortunately, each atom has a rather small probability $\eta$ to have such a spin exchanging collision during the relevant time $\tau$, namely
$\eta \approx \sigma v \tau \rho$, where $\sigma$ is the electron spin exchange
cross section  (for cesium it is $\sigma \approx 2\times 10^{-14}$ cm$^{2}$ \cite{Beverini71}), the density of atoms with opposite polarization is $\rho \approx 2.5\times 10^{10}$ cm$^{-3}$ and the speed is $v \approx 130$ m/s. 
During the interaction time $\tau=1$ ms the spin exchange collision probability
is thus $\eta \approx 6.5\times 10^{-3}$. The influence on the atomic quadratures can be found following the line in \cite{Madsen04}: the quadrature mean values are shortened by a fraction $\propto \eta$ and additional fluctuation  $\propto \eta$ is added to them. 
Since $\eta \ll 1$, states with phase-space features not finer than $\eta$ will not be seriously influenced by the spin exchanging collisions.



\section{Equal oscillation frequency}

In weak magnetic fields the oscillation frequency of coherences between
neighboring magnetic states is nearly proportional to $B$ and independent of
$m$. However, in stronger fields nonlinear Zeeman shift occurs that causes
differences between the oscillation
frequencies  of $\sigma_{-F,-F+1}$
and $\sigma_{F-1,F}$.
In particular, the difference between the neighboring
energy levels is up to the second order (see, e.g., 
\cite{JulsgaardThesis03,Julsgaard04})
\begin{eqnarray}
\Omega _Z(m) = \frac{E_{m+1}-E_m}{\hbar} =
\Omega_B - \frac{\Omega_B^2}{\Delta_{\rm HF}}(2m+1) ,
\label{ZeemanShift}
\end{eqnarray}
where $\Omega_B=\mu_B B/\hbar$ with $\mu_B$  the magnetic dipole moment and
$\Delta_{\rm HF}$ is the hyperfine splitting of the ground state, which for
cesium is $\Delta_{\rm HF} = 2\pi \times$9.19~GHz. The difference of rotation
frequencies of the two classes is $\Delta \Omega_Z 
= \Omega_Z(m=-F)-\Omega_Z(m=F-1) =
14\Omega_B^2/\Delta_{\rm HF}$. 
Assuming the typical light pulse duration $\tau
=$ 1~ms and oscillation frequency $\Omega_B \approx 2\pi\times 300$~kHz
as in \cite{FiurasekNature04,JulsgaardThesis03}, 
one
finds that the phase difference between the two coherences accumulated during
the evolution would be $\Delta \Omega_Z \tau \approx 0.3\pi$ which is not
negligible. 
In the double-cell scheme one can solve this problem by tuning the
magnetic fields in the two cells differently so that the field difference 
compensates for the quadratic Zeeman shift. This cannot be done in the single
cell scheme so that one has to look for another solution.

\subsection{Weak magnetic field}

One option is to work with weaker magnetic fields. For example,  with the
Larmor frequency $\Omega_B \approx 2\pi \times 50$~kHz the quadratic
Zeeman shift would lead to 
the accumulated phase difference of $\sim 20$~mrad which can be
neglected. However, the laser signal on the lower-frequency sidebands can be
much noisier than on higher frequencies so that one may prefer staying with
stronger magnetic fields.


\subsection{Ac Stark shift}

Another option is to use ac Stark shift compensating for the nonlinear Zeeman
shift. We propose using suitably 
polarized field  detuned from the D1 transitions, e.g. in cesium,
6$^2$S$_{1/2},F=4 \leftrightarrow$ 6$^2$P$_{1/2},F'=3$ and
6$^2$S$_{1/2},F=4 \leftrightarrow$ 6$^2$P$_{1/2},F'=4$.
Note that the signal close to the D2 transition does not interfere 
with this auxiliary
field.
As the best candidate appears  a field linearly polarized in the $x$ direction. 
The Stark shift of the magnetic levels
is then
\begin{eqnarray}
 E_S(m) = 
 \frac{I_S}{2\epsilon_0 \hbar c}
 \left(
 \frac{ |\mu_{m,m}^{(F'=3)}|^2}{ \Delta_{F'=3}}
 + \frac{ |\mu_{m,m}^{(F'=4)}|^2}{ \Delta_{F'=4}} 
 \right),
\end{eqnarray}
where the dipole moment squares are
\begin{eqnarray}
|\mu_{m,m}^{(F'=3)}|^2 &=& 
\frac{\epsilon_0 \hbar \lambda^3 \gamma}{2^7 \pi^2} (4-m)(4+m) , \\
 |\mu_{m,m}^{(F'=4)}|^2 &=&
\frac{\epsilon_0 \hbar \lambda^3 \gamma}{2^7 \pi^2} m^2 ,
\end{eqnarray}
$\Delta_{F'}$ is the detuning of the field with respect to the
transition to the
hyperfine level $F'$, and $I_S$ is the Stark-shifting field intensity.
If we denote by $\Delta_S$ the detuning of the field with respect to the
center of the two $F'=3,4$ transitions,
the shift of the frequencies 
between the neighboring levels 
$ \Omega _S(m) = [E_S(m+1)-E_S(m)]/\hbar$
is 
\begin{eqnarray}
\Omega _S(m) =
\frac{\lambda^3 \gamma I_S \Delta_2}{2^8 \pi^2 \hbar c}
\frac{2m+1}{\Delta_S^2 - \frac{\Delta_2^2}{4}} .
\label{StarkShiftPi}
\end{eqnarray}
Whereas for tuning between the two D1 transitions
to $F'=3,4$, $|\Delta_S| < \Delta_2/2$, 
the Stark
shift depends on $m$ in the same way as the 
quadratic Zeeman shift, for larger detunings
$|\Delta_S| > \Delta_2/2$ the two shifts 
(\ref{ZeemanShift}) and (\ref{StarkShiftPi}) have opposite dependences. 
In particular,
the $m$ dependent parts of the two shifts
cancel each other if the field intensity is 
\begin{eqnarray}
 I_S &=&  \frac{256 \pi^2 \hbar c  \Omega_B^2}{\lambda^3 \gamma 
 \Delta_{\rm HF}} \left( \frac{\Delta_S^2}{\Delta_2} - \frac{\Delta_2}{4}
 \right) .
\end{eqnarray}

When working with additional optical fields, one has to estimate their influence
on atomic decoherence by photon scattering.
The scattering rate
of photons absorbed and spontaneously reemitted by an atom
detuned by $\Delta'$ from the photon frequency
is \cite{JulsgaardThesis03} 
$\Gamma_{\rm ph} = \frac{\gamma}{2}\frac{s}{1+s},$
where the saturation parameter is
$ s=\frac{I_S}{I_{\rm sat}} \frac{1}{1+\left(\frac{2\Delta' }
{\gamma}\right)^2}, $
and the saturation intensity is
$I_{\rm sat}=\frac{2\pi^2\hbar c\gamma}{3\lambda^3}$.
We have to integrate  $\Gamma_{\rm ph}$ over $\Delta'$
with a  Gaussian distribution centered at
$\Delta_2/2$ and having the half-width $\Delta_{\rm Doppler}$. 
For our values 
with $\Omega_B \approx 2\pi\times 300$~kHz 
and assuming $\Delta_S=2\pi\times 3$~GHz,  the required intensity
is $I_S \approx 1$~mW/cm$^2$ and would lead to the scattering rate of 
$\Gamma_{\rm ph} \approx 18$~s$^{-1}$ which is
much less than the estimated
scattering rate due to the probe in the Copenhagen experiments
$\sim 130$~s$^{-1}$ \cite{JulsgaardThesis03}. This shows that the auxiliary
field will not disturb seriously the atomic memory.


\subsection{Ac Zeeman shift}

One can use microwave field tuned off-resonance with respect to the transition between the two hyperfine levels. Let us assume $\pi$-polarized field with magnetic field polarized in the $x$-direction,  whose frequency is detuned by $\Delta_{\mu}$ from the hyperfine frequency $\Delta_{\rm HF}$, 
and whose intensity is $I_{\mu}$.
The energy shift of the $m$ state is then 
\begin{eqnarray}
 E_{\mu}(m) = 
 \frac{I_{\mu}}{2\epsilon_0 \hbar c^3 \Delta_{\mu} }
  |\mu_{m,m}^{(\mu)}|^2,
\end{eqnarray}
where the magnetic dipole moment element between the $m$ states of the two hyperfine levels in cesium is 
$\mu_{m,m}^{(\mu)} = \mu_B \sqrt{1-(m/4)^2}$, 
and $\mu_B$ is the Bohr magneton. 
The shift of the frequencies 
between the neighboring levels 
$ \Omega _{\mu}(m) = [E_{\mu}(m+1)-E_{\mu}(m)]/\hbar$
is 
\begin{eqnarray}
\Omega _{\mu}(m) =
 \frac{I_{\mu} \mu_B^2}{32\epsilon_0 \hbar^2 c^3 \Delta_{\mu}}
(2m+1) .
\label{AcZeeman}
\end{eqnarray}
The ac Zeeman shift (\ref{AcZeeman}) compensates the nonlinear Zeeman shift
(\ref{ZeemanShift}) for the microwave field intensity
\begin{eqnarray}
 I_{\mu} = \frac{32 \epsilon_0 \hbar^2 c^3 \Delta_{\mu} \Omega_B^2}
 {\Delta_{\rm HF} \mu_B^2}.
\end{eqnarray}
If we use for the detuning $\Delta_{\mu}$ about ten times the frequency difference between the hyperfine transition with $m=-3$ and with $m=3$, i.e., 
$\Delta_{\mu}\approx 10\times 12\Omega_B = 2\pi \times 36$~MHz, the intensity with
$\Omega_B = 2\pi\times 300$~kHz should be
$I_{\mu}\approx 1.4$~W/cm$^2$.


\section{Quadrature rotations}

Although the above discussion shows the availability of the QND Hamiltonian
(\ref{QNDtot}), to be useful as a quantum memory medium, one has to be able to
manipulate to some extent the atomic degrees of freedom. An important
operation  is rotation of the atomic quadratures 
of each mode
\cite{Kuzmich00}, in our case,
e.g.,  $P_{A+} \to X_{A+}$ and  $X_{A+}\to -P_{A+}$,
and similarly for the ``$-$'' mode. In the two-cell schemes
this is achieved by magnetic pulses acting separately on each atomic sample.
For example, a magnetic $\pi/2$ pulse in one cell causes the Zeeman shift
yielding  the rotation  $P_{A1} \to X_{A1}$ and  $X_{A1}\to -P_{A1}$. However, 
{\em oppositely oriented} magnetic $\pi/2$ pulse in the other cell would cause
the corresponding rotation $P_{A2} \to X_{A2}$ and  $X_{A2}\to -P_{A2}$ which
would lead to the desired transformation of $X_{A+}$ and $P_{A+}$. The
definition of the collective quadratures $X_{A2}$ and $P_{A2}$ in (\ref{XA2})
and (\ref{PA2}) means that these quadratures would respond oppositely in weak
magnetic fields in comparison to $X_{A1}$ and $P_{A1}$, i.e.,
$P_{A2} \to -X_{A2}$ and  $X_{A2}\to P_{A2}$. Thus, if one  used the same
weak magnetic pulse for the two classes of atoms in the same cell, the two 
atomic modes would mix: $P_{A+} \to X_{A-}$, $X_{A+}\to -P_{A-}$, and
$P_{A-} \to X_{A+}$, $X_{A-}\to -P_{A+}$.

To solve this problem one can take advantage of the frequency difference of the
coherences $\sigma_{-F,-F+1}$ and  $\sigma_{F-1,F}$ caused by the  nonlinear
stationary
Zeeman effect, by the ac Stark shift, and/or by the ac Zeeman shift.

\subsection{Nonlinear Zeeman shift}

When using the nonlinear Zeeman
effect, one can apply a sufficiently strong magnetic field for a short time. 
During the magnetic pulse atoms  complete many 2$\pi$ rotations, however atoms
in the $m=-F$ class would end up half a rotation ahead in comparison to those in
the  $m=+F$ class.  To achieve this, one can use a magnetic pulse of duration
$\tau$ such that  $\Delta \Omega \tau = \pi$. If the time is chosen $\tau
\approx 30$~$\mu$s  (i.e., much shorter than the 1~ms reading and writing
pulses), one finds that the linear frequency should be $\Omega_B \approx
2\pi\times 3.1$~MHz corresponding to $B\approx 8.8$~G.   Although feasible in
principle,  it can be technically rather challenging to produce such strong,
precisely controlled, short magnetic pulses.

\subsection{Ac Stark shift}

Another option is to apply a Stark-shifting optical pulse, similarly as in the
preceding paragraphs.  The field produces state-dependent energy shift leading to
increased frequency difference  of the coherences $\sigma_{-F,-F+1}$ and 
$\sigma_{F-1,F}$.   When the pulse duration is 
$\tau$, the field intensity
should be such that
$\left| \Omega_S(m=F-1) - \Omega_S(m=-F) \right| \tau = \pi $.
Using Eq. (\ref{StarkShiftPi}) for a $\pi$ polarized pulse, 
one finds the condition
\begin{eqnarray}
 I_S = 
 \frac{32 \pi^3\hbar c }{7\lambda^3 \gamma \tau} 
 \frac{|\Delta_2^2-4 \Delta_S^2|}{\Delta_2} 
 \approx \frac{128 \pi^3\hbar c \Delta_S^2}
   {7\lambda^3 \gamma \Delta_2\tau}
 ,
 \label{ISpi}
\end{eqnarray}
where the approximation is valid for
 $|\Delta_S | \gg
\Delta_2/2$.
As an example, let us consider
$\Delta_S=2\pi\times 3$~GHz and $\tau=30~\mu$s, which would require 
 field intensity 
$I_S \approx$ 135 mW/cm$^2$. The number of scattered photons is in this
case  $n_{\rm phot}  \approx
0.06$ which is smaller than the number of photons scattered from the 
information carrying pulses $\sim 0.1$. 
Note that when increasing the detuning and field
intensity, the number of scattered photons approaches its limit
$n_{\rm phot} \to 24 \pi/7 \times \gamma/\Delta_2 \approx 0.04$.

\subsection{Ac Zeeman shift}

When a source of suitably polarized microwave field is available, one can 
take advantage of the $m$ dependence of the ac Zeeman shift as in Eq. (\ref{AcZeeman}). If a $\pi$ phase difference between the two atomic classes is to be achieved during time $\tau$, one needs the microwave field intensity
\begin{eqnarray}
 I_{\mu} = \frac{16\pi \epsilon_0 \hbar^2 c^3 \Delta_{\mu}}
{7 \mu_B^2 \tau}.
\end{eqnarray}
Assuming as in the preceding section $\Delta_{\mu} \approx 2\pi \times 36$~MHz
and $\tau = 30\ \mu$s we get the required  microwave field intensity
$I_{\mu}\approx$ 170~W/cm$^2$.


\section{Conclusion}

We have shown how a single atomic vapor cell can be used as a quantum memory for light with the quantum signal encoded at two sideband frequencies. The sideband encoding is required so as to suppress technical noises that would occur if one measures the light signal by integration of the cw homodyne signal over the pulse duration (typically a millisecond). The two sideband modes must be matched by two atomic modes; we have shown how to use two atomic coherences in one sample rather than using one cell for each atomic mode.

The single-cell approach  could substantially reduce the losses of light
occurring at the boundaries between different media (air, paraffin coated glass
walls of the cells, atomic vapor) and save the resources needed. The losses
would be especially disturbing if squeezed states are to be manipulated: 
absorption of  ${\cal A}$ at each boundary would add $\sim {\cal A}$ of vacuum
noise to the signal. Thus, with small losses ${\cal A}\ll 1$,  if a beam with a
perfectly squeezed quadrature crosses four boundaries (when working with the
double-cell scheme), the noise added to the squeezed quadrature would be twice as big as when crossing two boundaries (single cell scheme).

The scheme requires selective addressing of atoms in the $m=-F$ and $m=+F$
manifolds, which react differently to strong stationary magnetic fields (nonlinear Zeeman
effect), to Stark-shifting optical pulses, or to microwave fields (ac Zeeman shift).  Experimentalists will have to find the
best combination of these  approaches to trade-off between their  advantages
and disadvantages. When applied, the scheme should enhance our capabilities of
storing and processing quantum information carried by light.


\acknowledgments

I am grateful to J. Fiur\'{a}\v{s}ek, B. Julsgaard,
N. Korolkova, U. Leonhardt,
Yu. Rostovtsev,
J. Sherson, and E.S. Polzik for many stimulating discussions.
Very important inputs were suggested by the anonymous referee.
This work was supported 
by GA\v{C}R (202/05/0486), 
by M\v{S}MT (MSM6198959213),
by ESF (Short Visit Grant 647), 
and  by EU (QUACS RTN, contract No. HPRN-CT-2002-00309).



\begin{thebibliography}{99}
\bibitem{Beverini71}
N. Beverini, P. Minguzzi, and Strumia,
Phys. Rev. A {\bf 4,} 550 (1971). Note that this cross section was measured
by monitoring the decay of the spin in a spin polarized sample, and
may therefore not be directly applicable here. We can expect however, that
the relevant cross section is of similar magnitude.
\end{thebibliography}
\end{document}